# Quality-Quantity Trade-offs in Tests for Management of COVID-19-like Epidemics

Harish Sasikumar,[1] Manoj Varma,[1,2*]

1. Center for Nano Science and Engineering, Indian Institute of Science, Bangalore, 560012, India
2. Robert Bosch Center for Cyber Physical Systems, Indian Institute of Science, Bangalore, 560012, India
*Corresponding author: mvarma@iisc.ac.in

## Abstract

There are multiple testing methods to ascertain an infection in an individual and they vary in their performances, cost and delay. Unfortunately, better performing tests are sometimes costlier and time consuming and can only be done for a small fraction of the population. On the other hand, greater number of individuals can be tested using a cheaper, rapid test, but may only provide less reliable results. In this work, we studied the interplay between cost and delay of the tests as well the additional advantages offered by partial and complete lockdowns. To understand the influence of different test strategies, we implemented them on realistic random social networks with a COVID-19-like epidemic in progression. Specifically, we compared the performance of two tests mimicking the characteristics of popular tests implemented for COVID-19 detection. We present procedures and intuitive understanding to ascertain the optimum combination of the tests to minimize the peak infection as well as total quarantine days when the number of tests is constrained by a fixed total budget.

## Introduction

Non-pharmaceutical interventions (NPIs) which primarily restrict the movement of individuals and goods, are effective methods to mitigate and even contain epidemics. This has been evident from NPI strategies such as "informal quarantine policies by community leaders" during the flu outbreak (1918-19) [1] to the public health interventions during the SARS outbreak (2003) [2]. During the recent COVID-19 outbreak, it has been shown that in order to eliminate the epidemic from a population, four or more of these NPIs such as such as stay-at-home orders, limitations in gathering sizes and closing down of businesses and educational institutions have to be simultaneously implemented [3]. As NPIs are crucial in epidemic management, personal and community NPIs have been included in the strategic plans around the world to contain epidemics [4]–[6].

An ideal, complete lock-down, in principle, can quench any outbreak within the duration of a few infectious period. However, in most of the regions, complete lockdowns were either not implemented or not achieved to provide such quick results during the recent COVID-19 outbreak. Workarounds such as extended lockdowns, injudicious isolation or quarantining of individuals incur severe economic costs. Hence, scientific studies often advocate judicious usage of quarantining based on good testing strategies to minimize the economic impact [7].

An ideal testing and quarantining strategy minimizes the quarantine period along with preventing transmission events and outbreaks. The key aspect in an ideal strategy is to find as many infectious cases as possible in the first place. This is particularly challenging during Covid-19-like epidemics where a significant portion of the infectious population are asymptomatic. The initial studies suspect that the asymptomatic and pre-symptomatic infections can be as high as 81% [8]. This makes large scale testing crucial, along with NPIs. Large scale, randomized testing and subsequent quarantine-recommendations have already saved many lives. Even extremely large scale (of the order of the entire population) and frequent (even daily) testing strategies using unconventional methods are being considered [9].

As in any biological tests, testing for the presence of an infectious epidemic has multiple methods, each having their own advantages and disadvantages. In the context of Covid-19 outbreak as well, there are multiple diagnostic techniques [10]. There are accurate (high sensitivity and specificity) and costlier testing methods such as reverse transcription polymerase chain reaction (RTPCR), which typically have sample-to-result turnaround times from four hours to a few days [11]. There are also less accurate tests which are cheaper and faster with the results being provided within a few minutes [12]. In many places, it is often combinations of these tests that are implemented in the population [13].

In this work, we examine scenarios where a combination of tests, each with their own costs, performances and delays are employed and whose usage is constrained by a fixed total budget allocated for testing. We elucidate the combined effects of these tests in a network of individuals, of which we have no discernable indication of infection or recovery. This total ignorance makes all the individuals in the network to be equally likely to be infectious whereas only an unknown number of random individuals are infectious. This scenario is quite like the cases, for instance during the initial phase of an outbreak. Similarly, this scenario is also equivalent to a network of interconnected individuals who are the primary contacts of already detected infections.

To study the influence of different test strategies, we implement these strategies on random social networks with a COVID-19-like epidemic in progression. We begin our discussion with a deterministic compartmental model. This ascertains the basic characteristics of the progression of infection and testing methods. It also provides a system of differential equations onto which the effects of stochasticity are added. A more realistic model is developed by adding delay and stochasticity - both in transitions of individuals between compartments and in the results of the tests. The characteristics of real offline social network is implemented by creating networks based on a preferential attachment algorithm. The dynamics of test-based quarantining and restrictions in human interactions are studied by modulating the links between the nodes in the network. We consider two tests, one with high performance but with larger cost and slower turn-around times and the other which is rapid and cheap but with lower performance (reliability). We then investigate the optimum combination of these tests to suppress infection spread while reducing the detrimental effects of wrong quarantining.

## SITQR Model

We begin our discussions by proposing a modified continuous time deterministic compartmental model as in Figure 1. This susceptible-infectious-tested-quarantined (SITQR) model provides a framework to study the impact of testing procedures on the spread of infections. Our model is a direct extension of the classic susceptible-infected-recovered (SIR) model [14] with further fragmentations in the compartments as well as the effects of testing parameters taken into the model.

In this model, we subdivide the entire population into disjoint compartments shown as circles in Figure 1. Primarily, there are six compartments: unquarantined susceptible ($S$), unquarantined infectious (I) and unquarantined recovered ($R$) as well as their respective quarantined populations - $Q_S$, $Q_I$ and $Q_R$. In this context, quarantining (restrictions imposed on individuals suspected of infectiousness) is considered equivalent to isolation (restrictions imposed on individuals confirmed of infectiousness). We assume that the total population ($N$) as well as the number of individuals in each of these compartments is large so that they can be considered as continuous variables. In order to focus on the key aspect, i.e., the comparison of tests with varying performances, we neglect the vital dynamics. The key feature in our proposed model are the pathways from each of the unquarantined compartment to their respective quarantined compartment through testing procedures.

The standard assumptions in the spread of the epidemic are as following [15]:

1. From $S$ to $I$, the transition rate is assumed to be proportional to the fraction of unquarantined susceptible and infections with a proportionality constant $\beta$.
2. From $I$ to $R$, the transition rate is assumed to be proportional to the fraction of infectious individuals with a proportionality constant $\gamma$.
3. We assume that the duration of quarantine is designed according to the recovery rate so that the proportionality constant for de-quarantining is also $\gamma$.

For testing and quarantining, we make the following additions in the model.

4. We assume that a fraction of the total population ($\lambda$) is randomly selected for testing. Hence, the transitions from a compartment to its respective quarantined region, depends on $\lambda$ as well as the characteristics of the test.
5. We assume that the tests have the probabilities for identifying true positives and true negatives as $\eta^{TP}$ and $\eta^{TN}$, respectively. The probability of true positive determines the probability with which an infectious individual, when tested, is correctly identified, and is subsequently quarantined. It is the average sensitivity, when the test is employed on a large number of individuals [16]. Similarly, probability of true negative is defined as the probability with which an uninfectious individual (susceptible or recovered), when tested, is correctly identified, and is suggested not to be quarantined. It is the average specificity, when the test is employed on a large number of individuals [16].
6. Consequently, there are probability for false positive ($\eta^{FP}$) related to misidentifying an uninfectious individual to be quarantined as well as probability for false negative ($\eta^{FN}$) which misidentify an infectious individual and suggest not to be quarantined. They are complementary to $\eta^{TP}$ and $\eta^{TN}$ through the following equations

$$\eta^{FP} = 1 - \eta^{TN} \qquad\qquad 1$$

$$\eta^{FN} = 1 - \eta^{TP} \qquad\qquad 2$$

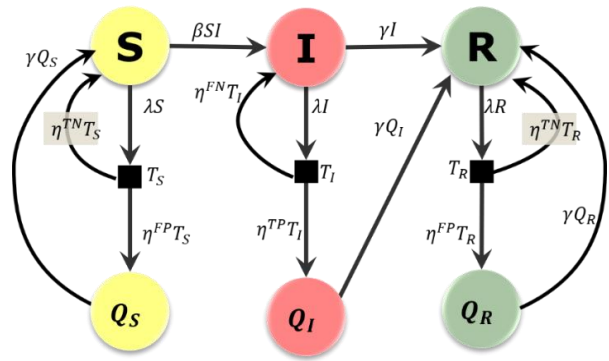

Figure 1. Flow diagram for the SITQR model with a single test. The six mutually exclusive classes are the unquarantined susceptible ($S$), unquarantined infectious ($I$), unquarantined recovered ($R$), quarantined susceptible ($Q_S$), quarantined infectious ($Q_I$) and quarantined recovered ($Q_R$). The tests are characterized by their probabilities to produce true positive ($\eta^{TP}$), true negative ($\eta^{TN}$), false positive ($\eta^{FP}$) and false negative ($\eta^{FN}$) results. $\beta$ is the factor deciding the infection rate and $\gamma$ is the factors deciding the recovery as well as unquarantining rates.

Further, the inherent assumptions in the model are

7. No births, deaths or travel into and out of the population. Hence, the total population ($N$) is constant and,

$$\dot{S} + \dot{I} + \dot{R} + \dot{Q_S} + \dot{Q_I} + \dot{Q_R} = 0 \qquad 3$$

where $\dot{X}$ denotes the derivative of $X$ with respect to time.

8. Latency period (time required for the infected to become infectious) is negligible.
9. For the time duration we are studying the system, the recovered population are immune to the infection and the immunity is permanent.

With the above assumptions, the dynamics of the epidemic model can be described by the following set of differential equations.

$$\dot{S} = -\beta SI - \lambda \eta^{FP} S + \gamma Q_S \qquad 4$$

$$\dot{I} = \beta SI - \lambda \eta^{TP} I - \gamma I \qquad 5$$

$$\dot{R} = \gamma(I + Q_I + Q_R) - \lambda \eta^{FP} R \qquad 6$$

$$\dot{Q_S} = \lambda \eta^{FP} S - \gamma Q_S \qquad 7$$

$$\dot{Q_I} = \lambda \eta^{TP} I - \gamma Q_I \qquad 8$$

$$\dot{Q_R} = \lambda \eta^{FP} R - \gamma Q_R \qquad 9$$

## Two-Test Model

The incorporation of a second test changes the compartmental model to a configuration as shown in Figure 2. We choose the fractions undergoing testing to be mutually exclusive. Hence, the total fraction of population being tested ($\lambda$) will be the sum of the fractions of the population being tested by the two separate tests.

$$\lambda = \lambda_1 + \lambda_2 \qquad 10$$

where $\lambda_1$ and $\lambda_2$ are the fraction of population tested with the first and the second test, respectively. In addition, the quarantining and de-quarantining procedures are the same for all the people who are tested positive and are independent of the test which identifies them as infectious. Hence, the only differences in the flow diagrams of the one-test and two-test models will be on the quarantining paths, from each unquarantined compartments to their respective quarantined counterparts.

We assume that the two tests employed have different true and false positive rates. Let $\eta_1^{TP}$ and $\eta_1^{FP}$ be the true and false positive probabilities, respectively, of test one. Similarly, $\eta_2^{TP}$ and $\eta_2^{FP}$ are the true and false positive probabilities for test two. In case where two tests are employed on two mutually exclusive, random fractions of the total population, all the earlier equations are valid with the following modifications.

$$\bar{\lambda}_{FP} = \lambda \eta^{FP} = \lambda_1 \eta_1^{FP} + \lambda_2 \eta_2^{FP} \qquad 11$$

$$\bar{\lambda}_{TP} = \lambda \eta^{TP} = \lambda_1 \eta_1^{TP} + \lambda_2 \eta_2^{TP} \qquad 12$$

These are obtained by equating the false positive fraction ($\bar{\lambda}_{FP}$) transitioning from the susceptible and recovered class to their respective quarantined class as well as true positive fraction ($\bar{\lambda}_{TP}$) transitioning from the unquarantined infectious class to the quarantined infectious class. Similarly, corresponding equations for the true negative and false negative fractions can also be written.

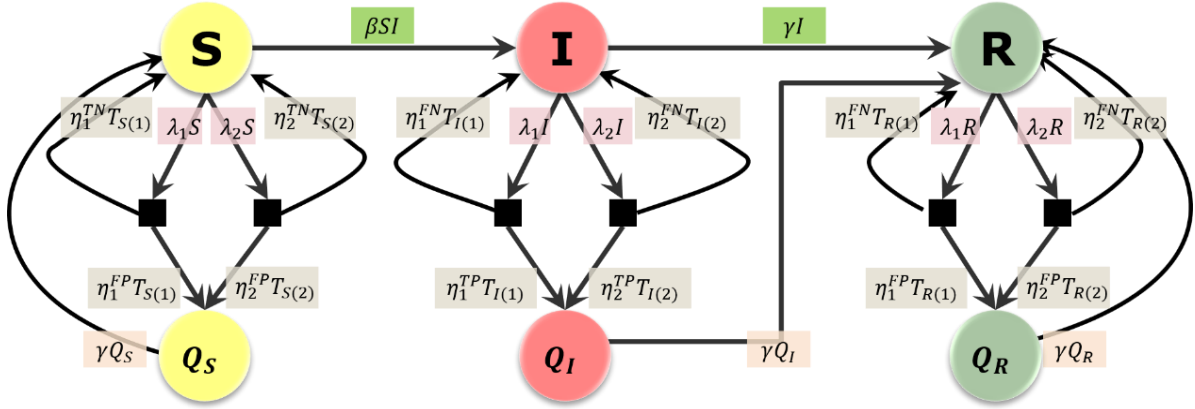

Figure 2. Flow diagram for the SITQR model with two tests. It contains the same classes as in Figure 1. The difference is in the transition of individuals from each of the unquarantined class to their respective quarantined classes. Here, it happens through two tests with fractions, $\lambda_1$ and $\lambda_2$ of the total population. $\eta_1^{TP}, \eta_1^{FP}, \eta_1^{TN}$ and $\eta_1^{FN}$ are the true positive, false positive, true negative and false negative probabilities, respectively, of test one. Similarly, $\eta_2^{TP}, \eta_2^{FP}, \eta_2^{TN}$ and $\eta_2^{FN}$ are probabilities for the second type of test.

## Cost Difference, Total Budget Constrain and Effective Performance

Without loss of generality, we assume that the second test is $m$ times cheaper than the first test. Hence, a total budget $B$, allocated for the entire testing procedure at any instance of time is split between the two tests as,

$$\text{B} = \lambda_1 + \frac{\lambda_2}{m} \; ; \begin{matrix} 0 \leq \lambda_1 \leq \text{B} \\ 0 \leq \lambda_2 \leq mB \end{matrix} \qquad 13$$

With this fixed budget assumption, $\lambda_1$ and $\lambda_2$ varies linearly and inversely with respect to each other (details in Section 1 in the Supplementary Information (SI)).

With the simplifying assumption that the falsely quarantined individuals do not aid (or hinder) the propagation of epidemic, the performance of a combination of tests can be quantified with its ability to detect the greatest number of infections. This is given as the average true positives ($\bar{\lambda}_{TP}$) as in Equation 12. With the fixed budget constrain (Equation 13), this is can be written as

$$\bar{\lambda}_{TP} = \lambda_1 \eta_1^{TP} + m(B - \lambda_1)\eta_2^{TP} \qquad 14$$

From this, the condition for $\bar{\lambda}_{TP}$ to increase with $\lambda_1 \left( \frac{d\bar{\lambda}_{TP}}{d\lambda_1} > 0 \right)$ can be found out as

$$m < \frac{\eta_1^{TP}}{\eta_2^{TP}} \qquad 15$$

It implies that $\bar{\lambda}_{TP}$ will be an increasing function of $\lambda_1$ only when the above inequality in satisfied. This can be seen in the variation of $\bar{\lambda}_{TP}$ as a function of $\lambda_1$ as in the few example plots given in Section 2 of SI. In short, a cheaper test is preferred over a costlier test if the cost of the cheaper test is lower than its reduced performance (true positive probability). These results are valid in a normally unrestricted society, i.e., a society where the individuals maintain their social physical contacts as usual, unless they are quarantined or isolated. This assumption of an unrestrictive society makes the untested population, essentially equivalent to the ones who are tested and found to be uninfectious. Thus, for the goodness of the society, a greater number of true positives must be isolated irrespective of the number of false negatives.

This comparison in the performance of tests is under many idealized assumptions. For instance, these results are independent of time. Hence, these comparisons are independent of the factors such as prevalence of the infection or quarantining. The impacts of these factors are not straight forward either. For instance, let us qualitatively consider the impact of quarantining. Despite the economic liability, quarantining of a true positive is preferred. However, the impact of quarantining of a false positive is not intuitive or even identical at different prevalence of the infection. On an individual level, a falsely quarantined individual isolates himself from the general population, thereby preventing himself from getting infectious. On the other hand, a false quarantine increases the density of infectious cases in the freely moving population and thereby enhance the disease spread. In any case, false quarantining always brings out unwarranted economic burden. Hence, to obtain a more realistic impact of test performance on the peak infection and total quarantines, a network model with stochasticity incorporated into the testing and epidemic spread is used.

# Epidemic Spread and Testing in Random Networks

The SITQR is a continuous time deterministic compartmental model governed by systems of differential equations and involves many idealizations. To study the propagation of epidemics and the influence of testing procedures as in real-world scenarios, network models are used [17]. These models avoid the assumptions such as homogeneous mixing. In addition, transmission of diseases, as well as testing can be modelled as stochastic events around their average characteristic parameters. In addition, in a network model, the spread of the disease depends on the number and pattern of contact among the individuals. Hence, the effects of social restrictions along with the random testing procedures can be studied by varying the average contacts between the individuals. By properly modelling the intervention procedures, they can be tested in these networks and thus bring out results that are closer to practical scenario. For instance, the events such as super-spreading will be inherently included in the model as a direct consequence of a highly connected infected node being not quarantined. These nodes directly infect a disproportionately large number of people and are pivotal during the early stages of the outbreak [18]. Similar direct and indirect consequences will be realistically mimicked in these simulations than in the deterministic model.

For our simulations, three sets of mutually independent parameters were used

1. Human network structure within the population
2. Parameters of the epidemic spread and recovery
3. Characteristics and extend of the random testing procedures

Each of these three sets of parameters were chosen such that they are comparable to the COVID-19 spread and testing procedures in real-world human networks. These three sets of parameters are explained in the following subsections.

## Offline Human Network

The individuals and their contacts were modelled as network ($G$) of nodes with links that interconnects them. The networks were implemented using igraph library in python [19]. $G$ was taken to be simple (without any loops or multi-edges) and undirected. Hence, $G$ can be represented as a pair of sets $G = (V, E)$ where $V$ is a non-empty set of vertices and $E$ is a set of 2-item subsets of $V$ called as edges or links [20]. Human social contacts form an approximately scale-free complex network with almost all nodes having an average number of connections and a few nodes with unusually large number of connections [21]. We made such a degree-distribution (distribution of number of edges connected to the nodes) among the nodes in our network using the preferential attachment algorithm. The resultant network mimics the characteristics of "small world networks" such as extremely small shortest paths among any two pair of nodes and power-law degree distribution [22]. The algorithm connects every newly formed node, preferentially to nodes which already have a greater number of links. For this, each of the newly formed node will have $\mu$ number of edges, connecting them to the i$^{th}$ preexisting nodes with a probability

$$p_i = \frac{d_i^k}{\sum_j d_j^k}$$

where $d_i$ is the degree of the i$^{th}$ node and the summation is made over all the preexisting nodes. $k$ is a positive exponent which decides the degree distribution. It can be varied in order to vary the distribution of links in the network. The effects of $\mu$ and $k$ on the structure of the network as well as the power-law degree distribution are shown in Section 3 of SI.

For the simulations, we took a population of constant size, $N = 1000$. Consequently, the order of the networks (number of nodes) was always 1000. Unless for simulations where $k$ and $\mu$ are varied they are taken to be 1 and 20, respectively. By taking $k$ to be 1, the preferential attachment produces the Barabási–Albert model which is linear and scale free and are quite close to real, large-scale networks [21]. $\mu$ was taken as a reasonable estimate from the most probable number of contacts of middle-aged individuals in an unrestricted social setting [23]. Consequently, the size of the network (number of edges) was $\approx$ 40,000. However this could change due to social restrictions and quarantining.

## Epidemic Spread and Recovery

Though the model presented in this work can be used for a wider array of diseases with varying infection and recovery processes, the results presented here are derived for an epidemic with characteristics similar to that of COVID-19.

We modelled the epidemic spread by associating one of the 6 states to each of the nodes. At each time instance, their states are updated by combination of stochastic events dictated by factor for infection ($\beta$), recovery ($\gamma$) and no change ($\alpha$). During each time step of the simulation, the event associated with a node is decided by a factor obtained by multiplying each of these factors with random numbers ($r_i$) drawn from a uniform distribution ($r_i \in X | X \sim U(0,1)$).

Let us assume $r_1, r_2$ and $r_3$ are three random numbers drawn from $X$. Then, the associated updation of the node is made depending on which of these factors dominate after the multiplication with the random numbers. For instance, when $\beta r_1 > \gamma r_2, \alpha r_3$, the node gets transitioned from a susceptible to an infected state. Similarly, in case of $\gamma r_2 > \beta r_1, \alpha r_3$ the node is transitioned from an infected to a recovered state and when $\alpha r_3 > \beta r_1, \gamma r_2$ the state of the node remains unchanged. These factors themselves are not constants and are kept closer to the real-world scenario by choosing their values appropriately.

The factor for infection ($\beta$) is defined in terms of an edge-induced subgraph ($G_I$) of $G$ which comprises of only the node under consideration and its immediate neighbors (as illustrated in Section 8 of SI). It is defined as

$$\beta = U\left(\frac{n_I}{n}\right)\beta_0 \qquad \text{17}$$

where $U$ is a unit step function whose value is 1 only when the node under consideration is susceptible and is 0 when it is either infected or recovered. This ensures that only a susceptible node gets infected. $n_I$ is the number of infected direct contacts of the node under consideration and $n$ is the total number of nodes in $G_I$. Hence, the terms inside the bracket determines the density of the infectious individuals in $G_I$. This is a modification from the SITQR model as local densities are considered here instead of the global average density. $\beta_0$ was adjusted to simulate the basic reproduction typically exhibited by COVID-19.

The definition of basic reproduction number ($R_0$) is the measure of the transmission potential of a disease, taking into account the contact rate in the population ($\mu$) as well as the nature of the infection [24]. It has been reported that $R_0$ of COVID-19 varies between 4.7 and 6.6. This implies that the number of infections double, approximately in every 2.4 days during the early stages of outbreak [25]. We varied the $\beta_0$ in our network to match this value, when no interventions are applied on the network. Accordingly, we found that the value of $\beta_0$ to be around 0.015 for a typical $\mu$ of 20 and is kept to be constant in all the simulations. A typical epidemic-curve when no interventions are applied, with an initial doubling period in between 2 and 3 days, is shown in section 4 of the SI.

The factor for recovery ($\gamma$) is designed so that the recovery of individuals from their infectiousness is similar to that from COVID-19. It has been reported that there are detectable viral loads from individuals even 10 days post-onset of the disease, especially in severe cases of infection [26]. However, there are no studies that detected live and infectious virus beyond day 9 [27]. This suggests that though the infection period of COVID-19 can be a few weeks or more, the infectious period is at most 9 days. Hence, we assumed that most of the individuals will get recovered from their infectiousness at 8$^{th}$ or 9$^{th}$ day post-onset. This kind of recovery was modelled using shifted and truncated Poisson function as following,

$$\gamma = \begin{cases} 0; \ t_n < \gamma_1 \\ \gamma_3(\gamma_2 - \gamma_1)^{(t_n - \gamma_1)}; \ t_n \geq \gamma_1 \end{cases} \qquad \text{18}$$

where $t_n$ is the number of days post-onset of the infectiousness. $\gamma_1$, $\gamma_2$ and $\gamma_3$ were adjusted to obtain most of the individuals to get recovered at the 8$^{th}$ or 9$^{th}$ day post-onset. They are nonunique fitting parameters and were chosen to be 7, 10 and 4 respectively to obtain a profile with a maximum of individuals getting recovered on the 8$^{th}$ day and the remaining few by the 9$^{th}$ day. Corresponding distribution of duration of infectiousness in a typical 1000-node network is given in section 5 of SI.

## Testing and Quarantining

In simulation, two different tests are employed on two mutually exclusive fractions of total population ($\lambda_1$ and $\lambda_2$), taken at random from the whole population. To study the effect of employment of one test over the other, the magnitudes of the fractions are varied systematically depending on the total budget ($B$). They vary linearly with each other and the slope depends on the relative costs of the tests ($m$) as described in Equation 13 as well as shown in section 1 of SI.

The process of testing is modelled as random experiments with conditional probabilities to report positive and negative test results, given that the tested node is infectious or uninfectious. Correspondingly, the test parameters - $\eta^{TP}$, $\eta^{TN}$, $\eta^{FP}$ and $\eta^{FN}$ are defined in terms of conditional probabilities and were given realistic numerical values. $\eta^{TP}$ is the conditional probability that the tests provides a positive result, given that the node is infectious. Complementary to this (equation 2), $\eta^{FN}$ is the probability that the test provides a negative result, given that the node is infectious. Similarly, $\eta^{TN}$ and $\eta^{FP}$ are respective probabilities for a negative and a positive test results, given that the node is uninfectious. They are complementary to each other as related by equation 1. Hence, for each test, there are two independent test parameters and they are chosen to be $\eta^{TP}$ and $\eta^{TN}$. For an average value from an infinite number of tests, these are respectively the sensitivity and specificity of the tests [16] and can be obtained from reported values in literature.

For simplicity, we assume that the tests are designed to report the infectiousness of nodes and that the infected patients are infectious as well. In the simulations we compare the effectiveness of RTPCR-like tests, which are in general accurate, costlier and time-consuming, against variants of rapid tests, which are less accurate, cheaper and provides test results practically instantaneously. We have taken RTPCR-like tests to be "test1" and a representative rapid test to be "test2". The sensitivity ($\eta_1^{TP}$) and specificity ($\eta_1^{TN}$) of test1 are taken to be 0.98 and 0.99 respectively, reasonable values reported for popular RTPCR tests [28]. Similarly, we chose the parameters of rapid tests to be conservative estimates from [12] with sensitivity ($\eta_2^{TP}$) and specificity ($\eta_2^{TN}$) 0.8 and 0.9, respectively. For comparing a more extreme case, we also included a less reliable rapid test with an $\eta_2'^{TP}$ as 0.50. Further we also assumed that the test2 does not have any delay between collecting the samples and providing the results, whereas test1 has a 1-day delay.

Irrespective of the tests or the states of the nodes, quarantine is employed for a fixed duration of 10 days on all nodes which are tested to be positive. This is in accordance with the recovery time and is assured to completely contain the period of infectiousness ($\leq 9$ days) [27]. There would be correct quarantining corresponding to nodes which are infectious and tested positive and wrong quarantining corresponding to nodes which are not infectious, but are tested to be positive. Due to fixed-duration of quarantining, there are also wrong quarantines when an infectious individual, who is tested positive remain in quarantine even after they are recovered. This can happen, for instance, when an infectious node is tested positive at the 4th day of its 9-day long infectious period. To study the effect of delay in obtaining the test results, the quarantining is implemented (for the positive test results) only after the respective delay of the tests. Accordingly, quarantine is implemented one day after a positive result from test1 where as it is implemented instantaneously for test2.

On a finite, small network as in this work, the epidemics which follow the SIR model get eliminated after limited duration from its onset. In order to faithfully compare performances of different test procedures, testing and quarantining must be avoided after the elimination of epidemic from the network. In real scenarios, we cannot be certain of this complete elimination within a short time duration and the epidemic could still reemerge from a few of the remaining unquarantined infected individuals. In such cases, the testing and quarantining must be restarted. In simulations, we established this using two thresholds based on the actual infections in the network. The first threshold ($I_{T1}$) determines the minimum number of infections in the network for the testing-quarantining procedure to begin or restart. Similarly, once the number of infections in the network is below the second threshold ($I_{T2}$), the testings will be stopped. The thresholds were set to be 1% and 0.5% which implies that the testing begins when the number of infectious cases in the 1000-node network is above 10 and stops when it is below 5.

## Implementation in Simulations

The networks were implemented using "igraph" library in python [19]. The updation of the states of each node at every time instance as well the diagnostic testing were implemented in the simulation using parallelized matrix multiplication algorithms in "numpy" [29] and pseudo-random number generators in "random" [30].

In all the cases, the simulations were run for 100 days with a time-resolution of 1 day. The simulations were carried out on 1000-node networks, with stochastic propagation and recovery of a Covid-19-like infection. As explained previously, the tests were also modelled stochastically and the nodes were quarantined for a fixed duration of 10 days, if found to be infectious. Two tests were employed on two mutually exclusive, random fractions of the total nodes, in such a way that the total budget spent is constant. Various parameters used to model distribution of contacts, spread of epidemic and the testing-quarantining strategy are all implemented as detailed in the previous sections. Summary of the parameters used in the simulation are given in Section 6 of SI and the flowchart of the simulation is presented in Section 7 of the SI.

For illustrative purposes the disease propagation, testing and quarantining in a smaller, 21-node network with $\mu = 10$ is shown in Figure 3. It shows the results from 4 different time instances during the progression of the epidemic. Each of the nodes (colored circles) indicate an individual and the lines indicate their interconnections. Note that the nodes could be susceptible (yellow), infected (red) or recovered (green) and their sizes are proportional to the number of connections. Figure 3(a) shows the initial condition. Figure 3(b) shows the network at the maximum prevalence of the epidemic with 10 nodes infected, 11 nodes susceptible and none recovered. Note that the nodes 3, 8, 10, 11, 12 and 14 are correctly quarantined whereas the nodes 15 and 18 are wrongly quarantined. Despite being susceptible, these nodes are quarantined due to false positives arising from the tests. Nodes 1, 5, 9 and 17 are not quarantined despite being infected. This could either be because they were not selected for the random tests or were reported to be (falsely) negative. Similarly, all the remaining nodes (0, 2, 4, 6, 7, 13, 16, 19 and 20) were either not tested or were tested and found to be negative. Figure 3(c) shows the situation nearing the end of the epidemic spread. Here nodes 3, 9 and 11 are recovered and are falsely quarantined. This could be due to the fixed-duration quarantine policy in our system where the infection ends typically in 8 or 9 days and the quarantine length is fixed to be 10 days. On the other hand, these could be the results of false positives. Figure 3(d) shows the network, immediately after the end of the epidemic.

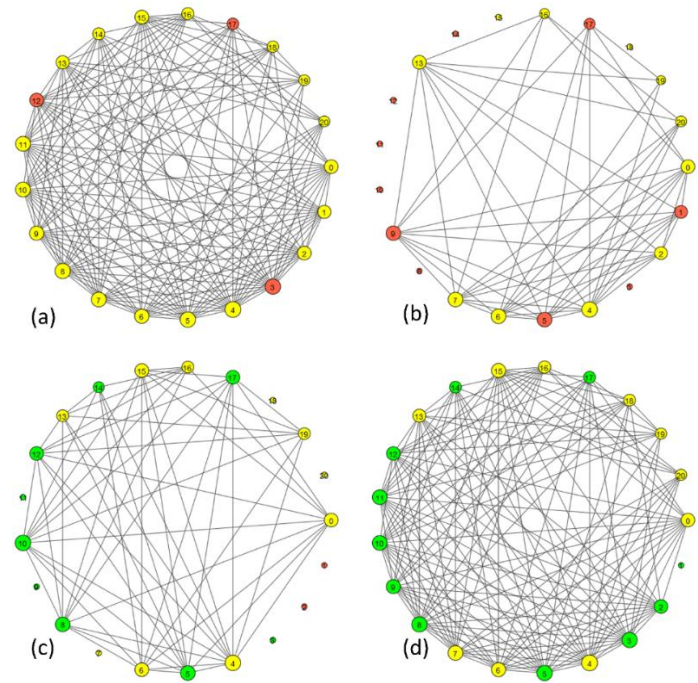

Figure 3. The propagation of epidemic in a network with 21 nodes. (a) Initial condition showing the default contacts when none of the nodes are quarantined (b) During the peak of the infection (c) Near to the end of the epidemic spread (d) Immediately after the end of the epidemic.

The actual simulations were done on 1000-node networks for a duration of 100 days. The output of each simulation has two kinds of results – the ground realities which cannot be known with certainty in real scenarios and the inferred results which are completely known. The results from the epidemic progression such as the number of susceptible, infected, and recovered as well as the classification of test results into number of true positives, false negatives, true negatives, and false positives are all ground realities which cannot be detected with certainty in real scenarios. On the other hand, the inferred quantities such as positive and negative test results are completely discernable in practice. In our simulations, both the ground realities as well as the test results could be stored and analyzed. In fact, in the later section we infer the goodness of a testing strategy based on a ground reality – peak infection – as well as on an inferred quantity – total quarantine days.

Figure 4 shows the simulation results of a 1000-node network with network, epidemic and simulation parameters as explained earlier (tabulated in Section 6 of SI). Additionally, each day, 67 of the unquarantined nodes are selected at random and are tested with RTPCR-like test ($\lambda_1 = 0.067$). Similarly, 33 nodes are tested with rapid tests ($\lambda_2 = 0.033$). Figure 4 is the "epidemic curve" of our proposed SITQR model showing the variation in the number of total susceptible ($S + Q_S$), infected ($I + Q_I$), tested ($T$), quarantined ($\Sigma Q = Q_S + Q_I + Q_R$) and recovered ($R$). As there are no births, deaths, immigration or emigration, the sum of all the mutually exclusive compartments (Sum $= S + I + R + Q_S + Q_I + Q_R$) remains to be $N$, all the time (as given in Equation 3). As the start and stop thresholds of the testing are 1% and 0.5% respectively, the testing started for $I + Q_I = 11$ and stopped for $I + Q_I = 4$.

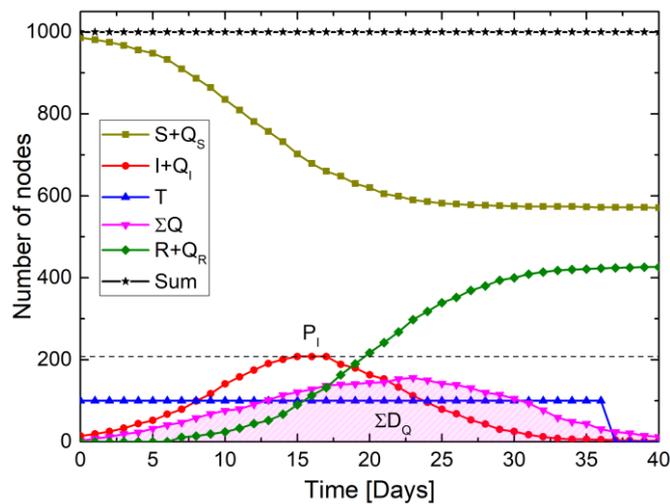

Figure 4. A typical epidemic curve as a COVID-19 like epidemic spreads through a 1000-node network (with parameters as given in Section 6 of SI) with testing and quarantining. Every day, 6.7% of the nodes are randomly tested with RTPC test with one day delay and a different 3.3% is tested with rapid test. The dashed black line indicates the peak infection ($P_I = 208$) and the shaded region under the total quarantine curve ($\Sigma Q$) indicates the total quarantine days ($\Sigma D_Q$).

Our simulations also revealed the performance of each of the tests in a daily basis and hence, their day-to-day sensitivity and selectivity. This indicates how each of the tests performs at different levels of prevalence of epidemic in the network. As input parameters for the simulations, the sensitivity and specificity of test1 were given as the probabilities for true positive ($\eta_1^{TP} = 0.98$) and true negative ($\eta_1^{TN} = 0.99$), respectively. Similarly, for test2 they were, $\eta_2^{TP} = 0.80$ and $\eta_2^{TP} = 0.90$, respectively. However, due to the stochastic nature of the tests, the calculated values of sensitivity and specificity of the tests varies from the ideal ones. From the simulations, the values for sensitivity and specificity can be calculated by applying their definitions [16]. The black curve in Figure 5(a) (with square markers) show the variation of true positives (TP1) out of 67 randomized RTPCR tests that are performed daily. Similarly, the blue curve shows that of rapid test (TP2). Their false negative counts are indicated as FN1 and FN2, respectively. Figure 5(b) shows the specificity values of the two tests (Spec1 and Spec2) and the associated data - the number of true negatives of test1 (TN1) and test2 (TN2) as well as false positives of test1 (FP1) and test2 (FP2).

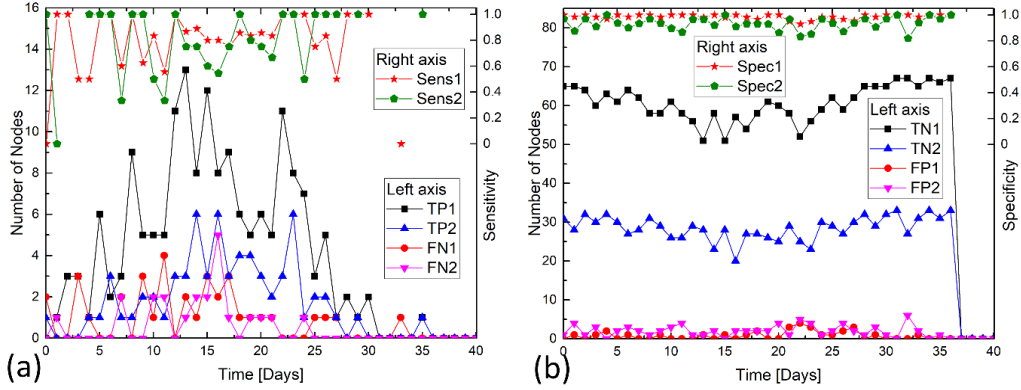

Figure 5. Variation of the number of daily (a) true positives (TP), false negatives(FN) and sensitivity (Sens). (b) true negatives (TN), false positives (FP) and specificity (Spec) of tests 1 and 2. These are the finer details from the simulation in Figure 4 where the infections are in the testing threshold in the first 37 days. Test1 with characteristics $\eta_1^{TP} = 0.98$, $\eta_1^{FN} = 0.02$, $\eta_1^{TN} = 0.99$ and $\eta_1^{FP} = 0.01$ was used daily on 67 random nodes. Test2 with characteristics $\eta_2^{TP} = 0.80$, $\eta_2^{FN} = 0.20$, $\eta_2^{TN} = 0.90$ and $\eta_2^{FP} = 0.10$ was done daily on a random but different 33 nodes.

## Evaluation of Testing Strategies

The allocation of budget to each of the tests were varied from the entire budget spending on test2 ($\lambda_1 = 0$, $\lambda_2 = mB$) to that on test1 ($\lambda_1 = B$, $\lambda_2 = 0$). The societal impact of these strategies on containing the epidemic is measured using the parameters peak infection ($I_P$) and total quarantine days ($\Sigma D_Q$) as well as a cost-function ($f$) which represented the cumulative societal impact. A typical curve indicating $I_P$ and $\Sigma D_Q$ is given in Figure 4.

Lowering the peak infection in the epidemic curve is crucial as it loads the healthcare system. Once the peak rises above the hospital capacity, mortality rate increases drastically due to insufficient healthcare facilities. Hence, the number of new infections must reduce with more and more rigor as we reach closer to this capacity. A judicious reduction of quarantining is also required as it reduces the social, psychological, and economic stress on the population.

Often reduction in peak infection and quarantining are favored by contradicting approaches. A complete lockdown keeps the peak infection to the minimum whereas the absence of any social restrictions eliminates all the burdens of quarantining. Ideally, we require a state where the restrictions are minimized without overloading the healthcare system. Hence, we require a trade-off which increasingly prefers quarantining of nodes over new infection as the number of infected cases approaches the hospital capacity. In such situations, typically power law utility functions are employed [31]. Hence, we introduced a similar utility function $f$ as

$$f = (I_P)^{\theta_1} + (\Sigma D_Q)^{\theta_2} \qquad \text{[19]}$$

The ranges of the exponents were fixed as $\theta_1 > 1$ and $0 < \theta_2 \leq 1$ to ensure that while minimizing $f$, quarantining is preferred in a non-linearly increasing fashion with increasing $I_P$. When $\theta_1 = 3$ and $\theta_2 = 1$, at the minimized value of $f$ and at low-prevalence (number of infected is 10) 30 quarantines or equivalently 300 total quarantine days are preferred over 1 infection. Whereas when the prevalence of the epidemic is higher (number of infected is 100), 300 quarantines are preferred over 1 infection. Further details and the procedure for choosing $\theta_1$ and $\theta_2$ are explained in Section 9 of the SI. While studying the impact of different testing procedures, the variation of utility function is also plotted in order to obtain the condition with judicious minimization of quarantining along with disease-control.

# Results

Simulations were done to study the effect of cost and delay of the tests as well as restrictions in social interactions for judiciously controlling the peak infection and quarantining. In our simulations, the allocation of the initial 11 cases of infections among the nodes, propagation of the epidemic as well as the testing procedures are all stochastic. Hence, exactly same initial conditions would result in different outcomes. In order to ensure the statistical significance of the results, averages were taken from 21 simulations for each of the data-point. Possible variability of the results is also indicated by whiskers which show the 80% confidence interval.

## Effect of delay in obtaining test results

A pronounced disadvantage of regular RTPCR tests is that their result turnaround time typically varies from a few hours to a few days. To study the effect of such delays, we tracked the modulation of peak infection ($I_P$) and total quarantine days ($\Sigma D_Q$) in our network after implementing RTPCR tests. Exclusively, RTPCR tests were done in the network with a budget, $B = \lambda_1 = 0.1$. Then we varied the turnaround times of RTPCR tests and positively tested nodes were quarantined immediately or with a delay of up to 4 days. The variation of peak infection and total quarantine days as a function of this delay is plotted in Figure 6.

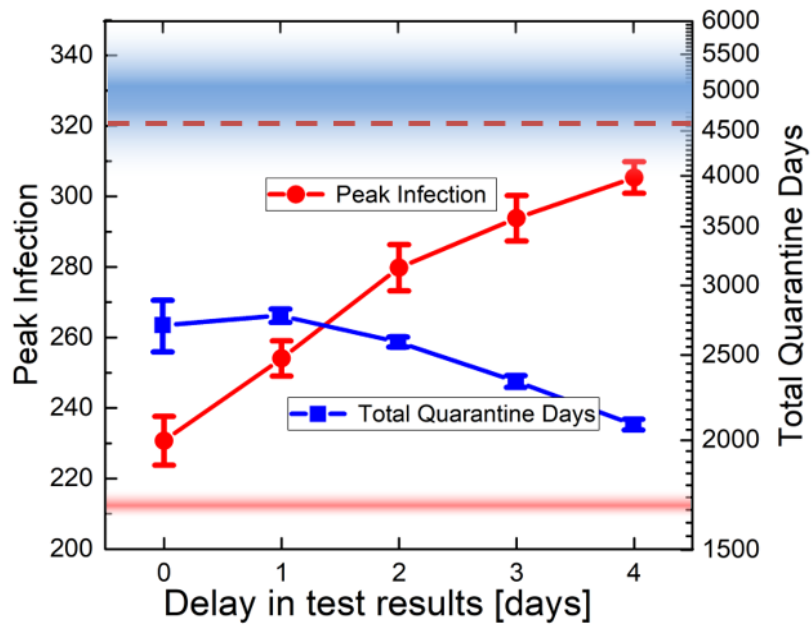

Figure 6. Variation of peak infection and total quarantine days as a function of delay in 1000-node scale free networks with $\mu = 20$. The network was tested using RTPCR with a budget, $B = \lambda_1 = 0.1$. The shaded regions represent peak infection (red) and total quarantine days (blue), when the same number of rapid-tests are employed in the network ($B = \lambda_2 = 0.1$). The dashed red line at $I_P \approx 320$ represents the average peak infection in a similar wild population (without any testing or quarantining). The peak infection increases by about 8.7% as the delay increases from zero to a single day. Similarly, the peak infection increases by about 5% as the RTPCR test delay increases from 3 days to 4 days.

For comparison, the average peak infection in a similar wild population (without any testing or quarantining) is plotted in red broken line ($I_P \approx 320$). For each day of delay, the peak infection increases by about 5% to 8.7%. There are two more aspects here which may be less apparent. They are listed and explained below.

1. The total quarantine days reduces with increasing turnaround delay of RTPCR. This is because delays expedites the infection as well as the disease-free equilibria in the network. As the tests are performed only when the number of infections in the network is within the threshold, this expediated disease-free equilibrium reduces the total number of tests and hence reduces the number of total quarantine days.

2. In terms of disease-control, rapid tests outperform hypothetical RTPCR tests which are delay-free. The red and blue shaded region represents the peak infection and total quarantine days, respectively, when only rapid-tests are employed in the network. Surprisingly, even with the same number of daily tests, less sensitive rapid tests were able to control the peak infection better than ideal delay-free RTPCR. The reason for this is the increased quarantining of the susceptible due the increased false positive probability of rapid test ($\eta_2^{FP} = 0.1$), compared to that of RTPCR ($\eta_1^{FP} = 0.01$). Similar results can be seen in the following section where, in terms of disease-control, rapid tests are shown to be preferred over RTPCR even when both the tests are of the same cost ($m = 1$).

## Effect of test cost

We studied how the reduction in the cost of the rapid test can affect the peak infection ($I_P$), total quarantine days ($\Sigma D_Q$) and subsequently, the utility function ($f$). Figure 7(a) and (b) shows the variation of the peak infection and total quarantine days as the test fraction of RTPCR changes from 0 to 0.1. Equivalently, the test fraction of the rapid test changes from a maximum value of $0.1m$ to zero. The red shaded region in Figure 7(a), near $I_P = 320$, is the expected peak infection in a similar 1000-node network without any testing or quarantining. It shows that, irrespective of the cost, all the testing strategies used in the simulations perform significantly better than the scenario without any testing.

Increasing the fraction of RTPCR testing reduces the peak infection only when the test is cheaper than the rapid tests ($m = 0.5$). Even with that, the peak infection is close to the no-test scenario. In all other cases, the peak infection increases as the number of RTPCR test increases. Interestingly, even for m=1, the condition where RTPCR tests are as cheap as rapid tests, rapid tests are preferred. This is primarily because, though the rapid tests can detect only a fewer true positive ($\eta_2^{TP} = 0.8$) compared to RTPCR tests ($\eta_1^{TP} = 0.98$), it quarantines many more of the false positives ($\eta_2^{FP} = 0.1$ $\eta_1^{FP} = 0.01$). It is directly evident from m=1 curve in Figure 7(b) where the total quarantine days is larger ($\Sigma D_Q \approx 4200$) when there are only rapid tests ($\lambda_1 = 0$), compared to a situation when there are only RTPCR tests ($\Sigma D_Q \approx 2800$ at $\lambda_1 = 0.1$).

The wrongly quarantined nodes are equivalent to self-isolated or reverse quarantined individuals. If we overlook the increased quarantine and its associated economic implications, these measures can reduce the infection rate and peak infection. Thus, solely in terms of disease-control, a test which misidentifies susceptible to be infectious is preferred. This result points to the significance of self-isolation or reverse quarantining for reducing the peak infection. Another reason for the rapid tests to outperform RTPCR at m=1 is the one-day result turn-around delay of the RTPCR tests. But from Figure 6(a), it is apparent that even with 0-day delay for RTPCR, the peak infection averages only around 230, if only RTPCR tests are used ($\lambda_1 = 0.1$). This implies that the removal of the one-day delay in RTPCR test can reduce the amplitude, but cannot change the sign of the average slope of m=1 curve in Figure 7(a). Even the degraded test 2, which has its sensitivity among the least of the practical diagnostic tests ($\eta_2'^{TP} = 0.50$), performs similarly to the normal rapid tests in terms of disease control.

However, this is enabled by quarantining a significantly larger number of people ($\Sigma D_Q \approx 9800$ from Figure 7(b)).

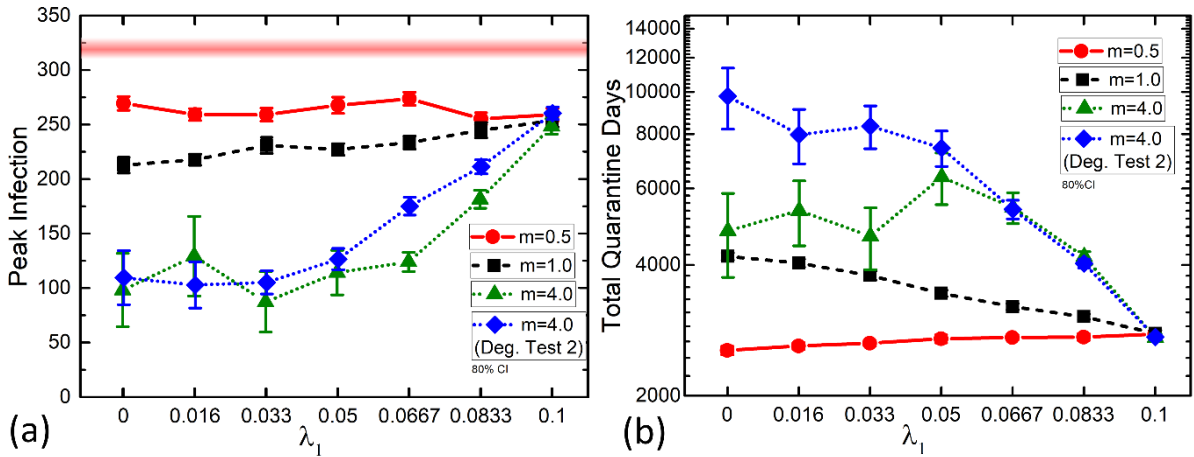

Figure 7. Effect of the cost of rapid-test relative to RTPCR. Variation of (a) Peak infection and (b) Total quarantine days as the ratio of the fraction of population tested with RTPCR test ($\lambda_1$) varies.

Hence as an oversimplified remark specific to disease-control, we can say that the quantity of testing and widespread quarantining is preferred over the quality of the tests. However, in case of inferior tests, it is the quarantining, rather than testing which controls the peak. As injudicious quarantining is not advisable, the utility function is plotted (Figure 8) to obtain a better trade-off between peak infection and total quarantine days.

The interplay between peak infection and total quarantine days makes the variation of $f$ non-monotonous for m=4 scenarios as shown in Figure 8. For regular rapid tests with m=4, it reaches the minimum at $\lambda_1 = 0.062$. This means that in the 1000-node network, everyday around 62 have to be tested with RTPCR and with the remaining budget, 152 rapid tests need to be performed. This is an example case and the ratio will be different for tests with different parameters. Using the same procedure, optimum test-fractions can be derived for tests with other performance parameters as well.

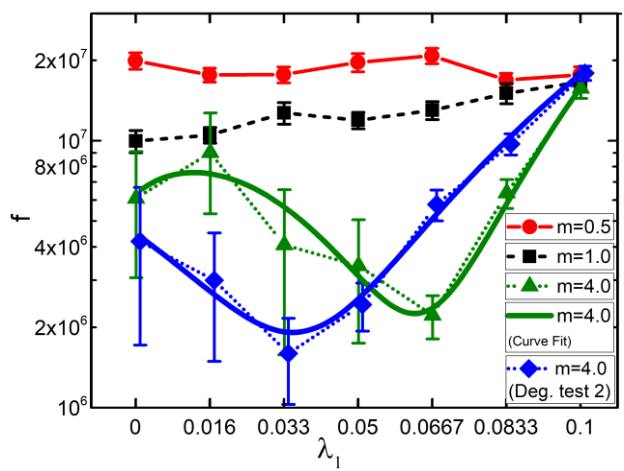

Figure 8. Variation of the utility function, $f$ (as in Equation 19) as the ratio of the fraction of population tested with RTPCR test ($\lambda_1$) varies. The function has to be minimized for the optimal combination of peak infection and total quarantine days. Here, the minimum is at $\lambda_1 = 0.062$ as RTPCR and regular rapid tests of $m = 4$ are done in the network. For the RTPCR - degraded rapid test combination, the minimum is at $\lambda_1 = 0.034$

The conditions discussed so far considered the effect of quality, cost, and the turn-around time of the tests. They suggested the combination of tests required to manage the epidemic spread with reasonable trade-off between peak infection and total quarantine days.

## Effect of restricted social interactions and complete lockdown

We have seen that the crucial factors preferring rapid tests are their reachability and thus its ability to provide restriction in social interactions over a wider population. In this section we study the combined effects of restricted social interactions along with testing-quarantining strategies. Social restrictions reduce the average contacts between the nodes. This can be simulated in the network by reducing the average number of edges per new nodes ($\mu$) while creating the network using the preferential-attachment algorithm. In the previous simulations, we assumed that in the unrestricted society, $\mu$ is 20. Here, we evaluate the performance of various testing strategies in societies where the average social contacts differ. Societies are grouped into ones with high social contacts ($\mu = 30$), regular ($\mu = 20$), partially restricted ($\mu = 10$) and under lockdown ($\mu = 2$). Figure 9(a) and (b) shows the variation in peak infection and total quarantine days, respectively, when the testing fraction of RTPCR and rapid test (with $m = 4$) is varied. Even by reducing the per-capita contact by half ($\mu = 20$ to $\mu = 10$), the peak infection gets reduced to 22% and 43%, respectively, in the best and the worst cases. Also, the complete lock-down ($\mu = 2$) quickly removes the epidemic from the network as can be seen by the peak infection remaining very close to the initial infection count of 11.

From the positive slopes of the curves in Figure 9(a) it is apparent that rapid tests are preferred in terms of disease-control. However, the preference becomes less significant as the social restrictions increases. Scenarios with regular and high social contacts have average slopes of $1.4 \times 10^3$ and $2.7 \times 10^3$ $P_I$ per $\lambda_1$, respectively, showing significant increase in peak infection as the fraction of RTPCR increases. During restricted and lock-down conditions, these slopes reduce by orders of magnitudes to $3.7 \times 10^2$ and $2.3 \times 10^1$ $P_I$ per $\lambda_1$, respectively.

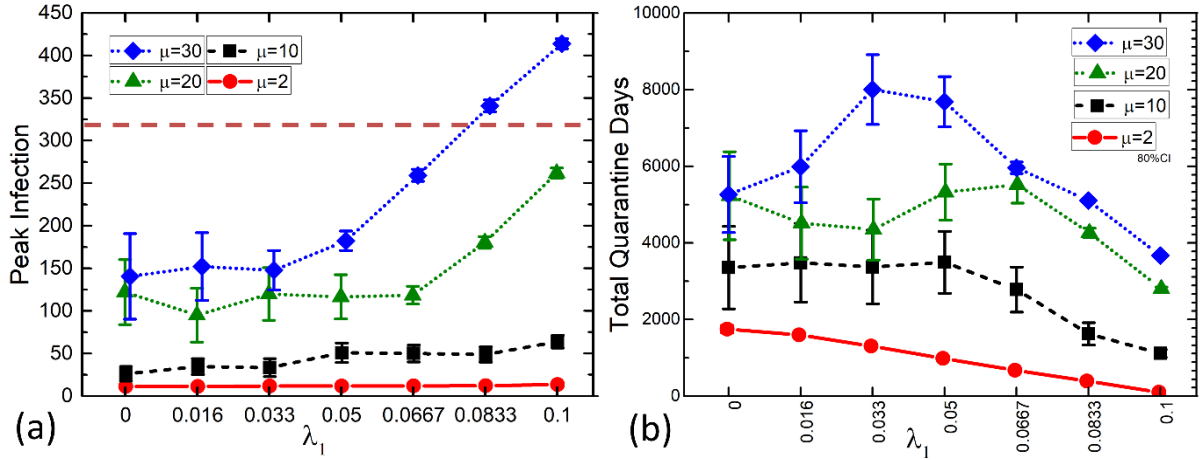

Figure 9. Effect of introduction of restricted social interactions and complete lockdown along with the combination of RTPCR and rapid tests. The total budget, $B = 0.1$ and rapid tests are four times cheaper ($m = 4$) than RTPCR. Social restrictions are simulated in the network by reducing the average number of edges per new nodes ($\mu$). The conditions are - high social contacts ($\mu = 30$), regular ($\mu = 20$), partially restricted ($\mu = 10$) and under lockdown ($\mu = 2$). (a) Variation of peak infection (b) Variation of total quarantine days.

The variation of total quarantine days in these scenarios are given in Figure 9(b) and the variation of the utility function is given in Figure 10. As there are practically no new infections in the complete lock-down scenario ($\mu = 2$), the testing and quarantining can very well be avoided in this situation.

An interesting finding from these simulations is that the optimal test combination depends on density of human contacts as well. From Figure 10, it is apparent that a network with only 50% of the normal contacts ($\mu = 10$), the entire budget must be spent on the rapid tests as the minimum of $f$ is obtained for $\lambda_1 = 0$. Under similar budget and test costs, the fraction of the RTPCR required for regular ($\mu = 20$) and high contact ($\mu = 30$) scenarios are $\lambda_1 = 0.067$ and $\lambda_1 = 0.033$, respectively.

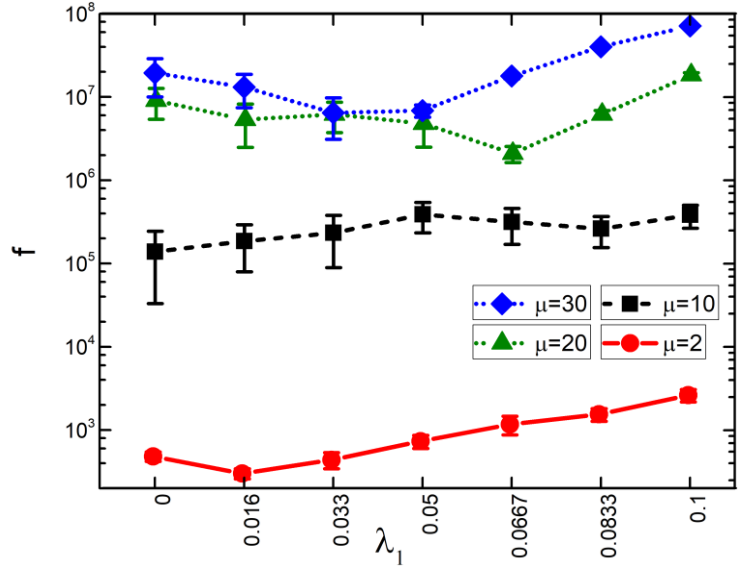

Figure 10. Variation of the utility function, $f$ (Equation 19) under various intensities of social interactions and combination of tests as in Figure 9.

By choosing the optimum test combinations, the peak infection (from the untested scenario with $I_P \approx 320$) can be reduced to 31.3%. The reduction of the per-capita contacts by half ($\mu = 20$ to $\mu = 10$) could significantly reduce it to a final value of around 16% of the untested scenario.

## Conclusion

In this work, we examined scenarios where tests are performed on random individuals using combinations of RTPCR tests and rapid tests. The social interactions were modelled by scale-free social networks on which a COVID-19 like epidemic is in progression. Though the properties of the epidemic spread are made closer to the COVID-19 epidemic, there were some compromises with the actual scenario to maintain the simplicity and focus on crucial results. This has led to a few caveats and the following are the major three.

1. Immunity of the recovered population is not permanent. The natural immunity against COVID-19 is not long lasting especially in persons with mild symptoms [32]. Moreover, there are vertical transmissions [33] as well as maternal passive immunity [34].
2. All infected people are not infectious. The infectious period of COVID-19 ($\leq 9$ days) is shorter than the period in which they are found to be infected [26].
3. The population in a network are more dynamic than our model. The additional dynamics typically include births, deaths, immigration and emigration.

In this work, we started from a deterministic compartmental model and an associated system of differential equations onto which the effects of stochasticity were added. The dynamics of test-based quarantining and restrictions in human interactions were studied. We elucidated a procedure to understand the trade-off between cost and performance of tests under the constrain of a total budget allocation.

We showed how the characteristics of the tests impacts the peak infection and total quarantine days and how to judiciously minimize them using a utility function. We have seen that the increase in result turnaround time increases peak infection by 5% to 8.7% for each day of additional delay. In terms of disease-control, quantity of testing and widespread quarantining is preferred over the quality of the tests. However, in case of inferior tests, the increased false quarantining plays a crucial role in controlling the peak infection.

We have shown that for an optimal reduction of peak infection and total quarantine days using a combination of tests with different characteristics, there is an ideal ratio. In case of our 1000-node scale-free networks with COIVD-19 like epidemic in progression, 62% of the budget must be spent on RTPCR and the remaining on rapid tests, when rapid tests are 4 times cheaper. The same procedure can be utilized to find the optimum for other diseases and test-combinations. Finally, we studied how the reduction of human contacts perform in comparison with the optimum testing strategies.

This work provides an intuitive understanding about the relationship between cost, accuracy, and delay in obtaining the result of the tests and how well they fair in comparison with complete lockdown and conditions with partial restrictions. This understanding is important to decide the quality and quantity of tests and thereby to implement better strategies for nonpharmaceutical interventions.

# Supplementary Information:

## Quality-Quantity Trade-offs in Tests for Management of COVID-19-like Epidemics

Harish Sasikumar,[1] Manoj Varma,[1,2*]

1Center for Nano Science and Engineering, Indian Institute of Science, Bangalore, 560012, India
2Robert Bosch Center for Cyber Physical Systems, Indian Institute of Science, Bangalore, 560012, India
*Corresponding author: mvarma@iisc.ac.in

## 1. Fixed Budget Constrain

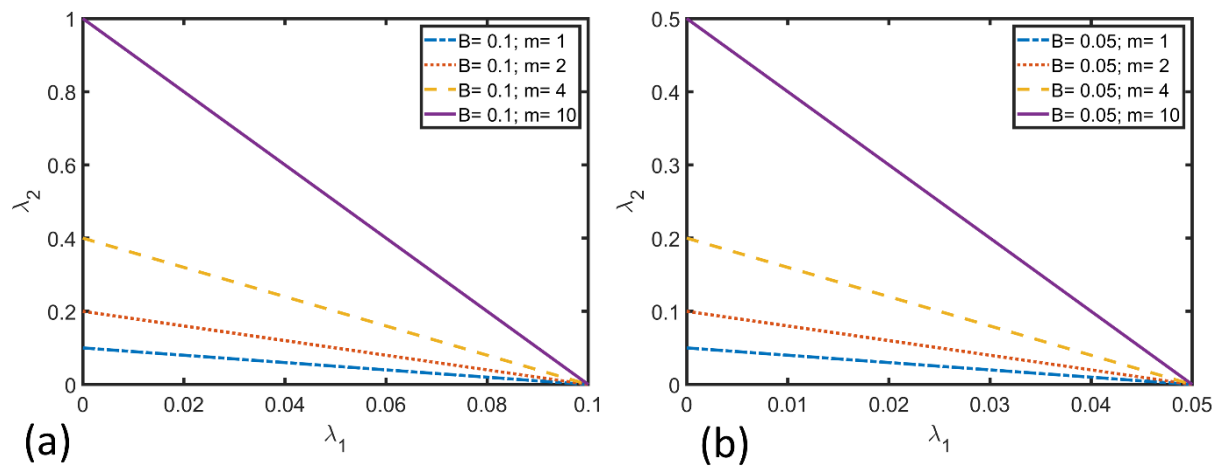

SI Figure 1. Variation of fraction of total population tested with first test ($\lambda_1$) and second test ($\lambda_2$) for a fixed budget, B and when the second test is m times cheaper than the first test. $\lambda_1$ and $\lambda_2$ varies inversely and linearly with each other. (a) $B = 0,1$. Accordingly, the range of $\lambda_1$ is $[0, 0.1]$, whereas that of $\lambda_2$ is $[0, 0.1m]$. (b) B=0.05. Ranges of $\lambda_1$ and $\lambda_2$ are $[0, 0.05]$ and $[0, 0.05m]$, respectively.

# 2. Effect of the Cost of the Tests

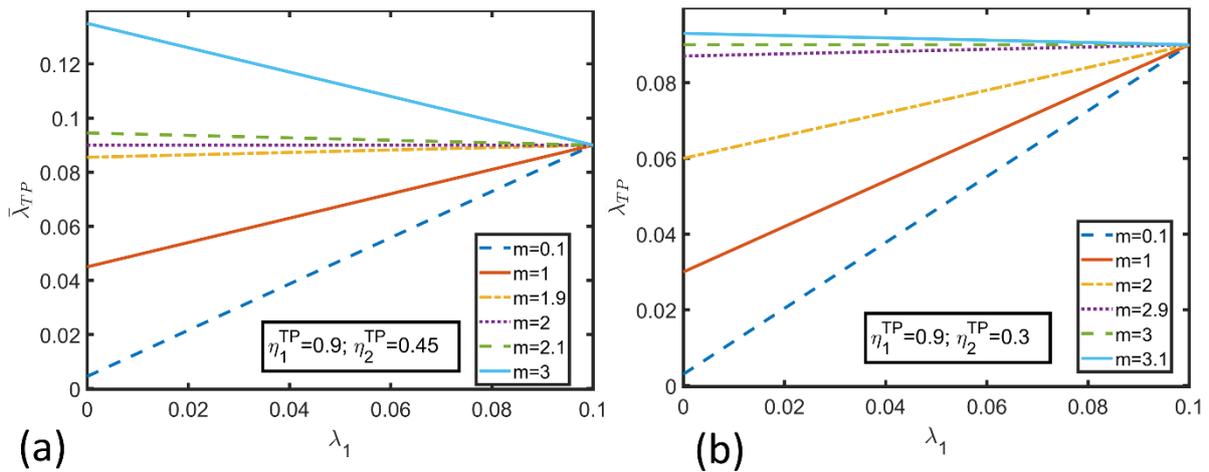

(a)

(b)

SI Figure 2. Variation of $\bar{\lambda}_{TP}$ as a function of $\lambda_1$. (a) $\eta_1^{TP} = 0.9$ and $\eta_2^{TP} = 0.45$ such that $\frac{\eta_1^{TP}}{\eta_2^{TP}} = 2$. Accordingly, $\bar{\lambda}_{TP}$ increases with $\lambda_1$ for $m < 2$ and reduces with $\lambda_1$ for $m>2$. (b) Similarly, with $\eta_1^{TP} = 0.9$ and $\eta_2^{TP} = 0.3$ where the transition of $\bar{\lambda}_{TP}$ from an increasing to a decreasing function with respect to $\lambda_1$ happens at $m = 3$.

# 3. Structure of the Offline Social Network

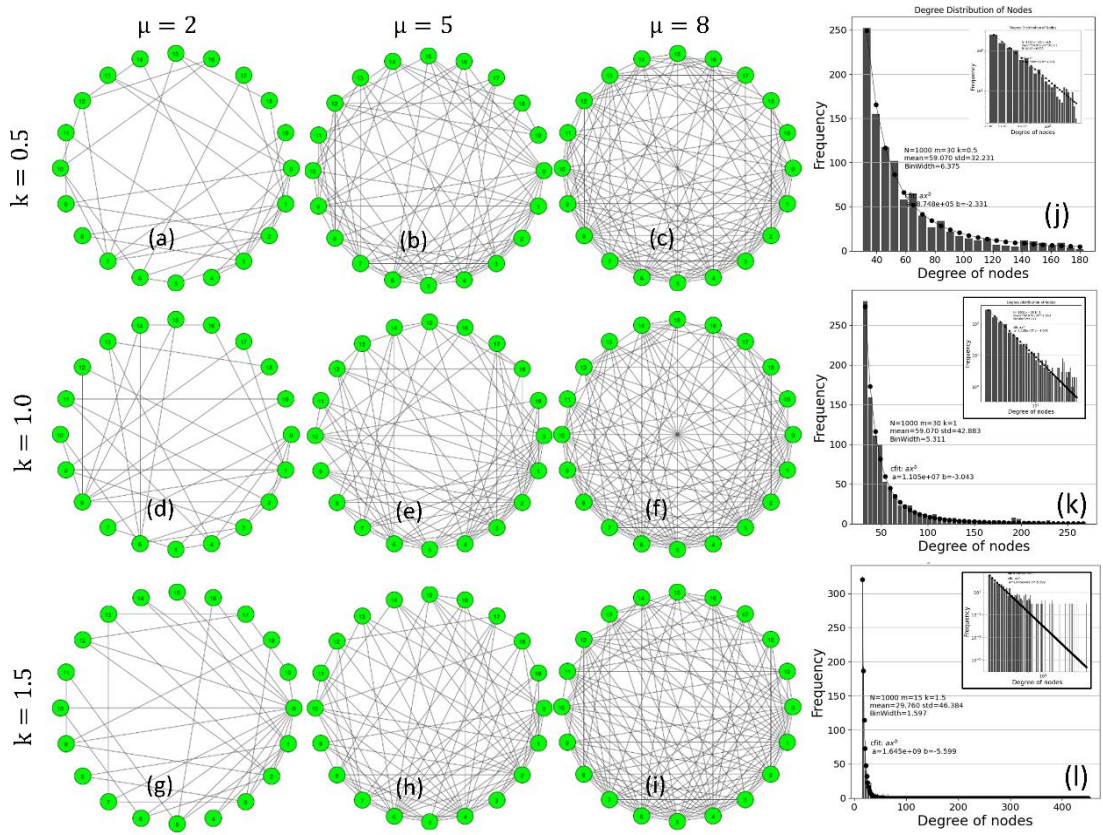

SI Figure 3.

Relation between k (NLPA) and a (exponent in degree distribution), Relatin between mu and mean, Extended exponential. Long tale, Exponential

# 4. Modelling the Probability of Infection

Without any testing or quarantining ($B = 0$) and for $\gamma_1, \gamma_2, \gamma_3, \alpha$ and $\mu$ as in SI Table 1, $\beta_0$ was varied until the initial epidemic doubling period is in between 2 and 3 days. SI Figure 4 shows a result with such a doubling period. The number of infections in days 0, 1, 2 and 3 since the outbreak of epidemic in the network are 14, 19, 24 and 40, respectively. For this, $\beta_0$ was found to be around 0.015.

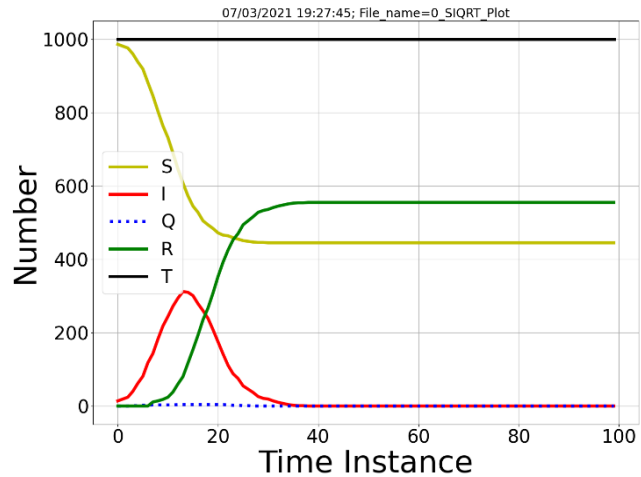

SI Figure 4. Epidemic curve with no intervention (B=0) and epidemic and network parameters as given in SI Table 1. The number of infections in days 0, 1, 2 and 3 since the outbreak of epidemic in the network are 14, 19, 24 and 40, respectively. This makes the initial doubling period to be in between 2 and 3 days. Note that the peak infection ($I_P$) is 312 and a total of 555 nodes were infected.

# 5. Modelling the Probability of Recovery

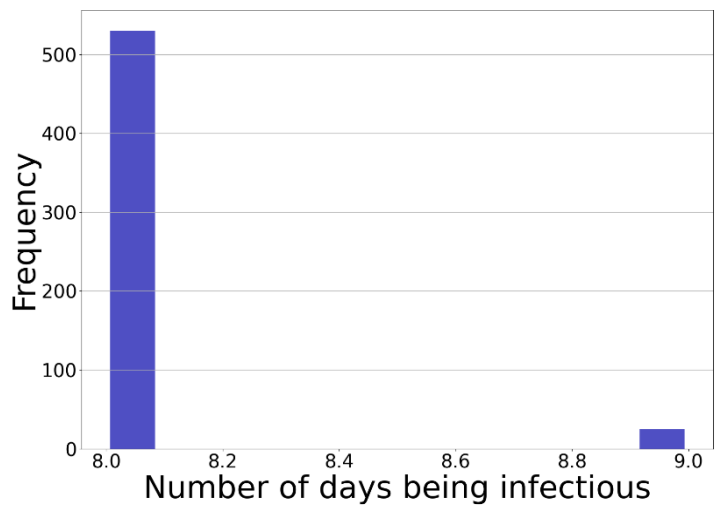

SI Figure 5. Distribution of number of days required for recovery for the 555 infected nodes in the simulation from which SI Figure 4 is plotted. Note that the shape of this distribution remains unchanged irrespective of the employment of non-pharmaceutical interventions.

# 6. List of Parameters and their Nominal Values

SI Table 1.

| Type | Parameter | Value | Notes | Ref. |
|------|-----------|-------|-------|------|
| Network | $N$ | 1000 | Population. Number of nodes in the network. | |
| | $k$ | 1 | Coefficient for the preferential attachment in the formation of network | [21] |
| | $\mu$ | 20 | Average number of contacts | [23] |
| Epidemic (Infection and Recovery) | $\beta_0$ | 0.015 | Fraction proportional to the probability that an infectious individual infects one susceptible in a single day when it is connected to 30 susceptibles. | [25] |
| | $\gamma_1$ | 7 | Factors deciding the fraction proportional to the probability that an infectious individual is recovered in the $t^{th}$th day post-infection. $$\gamma = \begin{cases} 0; & t_n < \gamma_1 \\ \gamma_3(\gamma_2 - \gamma_1)^{(t_n - \gamma_1)}; & t_n \geq \gamma_1 \end{cases}$$ $t_n$: number of days post-onset of the infectiousness. $\gamma_1, \gamma_2, \gamma_3$: parameters adjusted to obtain most of the individuals to get recovered at $8^{th}$ or $9^{th}$ day post-onset. | [27] |
| | $\gamma_2$ | 10 | | |
| | $\gamma_3$ | 4 | | |
| | $\alpha$ | 1 | Fraction proportional to the probability that no change happens to the node. | |
| | $Q_D$ | 10 | Number of quarantine days after the nodes were teste positive. Based on the maximum days for which the nodes are infectious. | [27] |
| | $I_0$ | 11 | Initial number of infected individuals. | |
| | $S_0$ | 989 | $$N - I_0$$ Initial number of susceptible individuals. There were no recovered of quarantined individuals at the beginning of the simulations. | |
| Budgeting (and Utility Function) | $B$ | 0.1 | Total budget allocated. From Equation 13 $$B = \lambda_1 + \frac{\lambda_2}{m}; \quad 0 \leq \lambda_1 \leq B; \quad 0 \leq \lambda_2 \leq mB$$ This implies that the 10% of the entire population can be tested every day if first test is employed. | |
| | $m$ | [0.5, 4] | How many times, the second (rapid-like) test is cheaper than the first (RTPCR-like) test. Values below 1 are typically not observed and are plotted for curiosity. Values above 4 are possible. | |
| | $\lambda_1$ | $0 \leq \lambda_1 \leq 0.1$ | Fraction of total population that is tested with the first (RTPCR-like) test. | |
| | $\lambda_2$ | $0 \leq \lambda_2 \leq 0.1m$ | Fraction of total population that is tested with the second (rapid-like) test. As the second test becomes cheaper (as $m$ becomes larger), larger fraction of the total population can be tested if test 2 is employed | |
| | $t = \frac{\theta_1}{\theta_2}$ | 3 | Factor deciding how much of quarantine is preferred over an infection. | |
| Simula | $t$ | 100 days | Total duration of the simulation. However, the testing and quarantining will not be performed in all of these days. | |

| | | | Testing happens according to the threshold values for the number of infections as given in the next section. | |
|---|---|---|---|---|
| | $\Delta t$ | 1 day | Time-resolution of the simulation. The updation of epidemic spread, testing and quarantining is done on a daily basis in the simulation. | |
| | $r$ | 21 | Number of times the simulation is run to obtain the | |

## Parameters of Tests and Quarantine

SI Table 2. Parameters specific to tests.

| | **Test 1** (similar to RTPCR) | **Test 2** (similar to Rapid tests) | **Degraded Test 2** |
|---|---|---|---|
| **TP** | $\eta_1^{TP} = 0.98$ | $\eta_2^{TP} = 0.80$ | $\eta'^{TP}_2 = 0.50$ |
| **TN** | $\eta_1^{TN} = 0.99$ | $\eta_2^{TN} = 0.90$ | $\eta'^{TN}_2 = 0.90$ |
| **Delay** | 1 day | 0 day | 0 day |
| **FN** | $\eta_1^{FN} = 0.02$ | $\eta_2^{FN} = 0.20$ | $\eta'^{FN}_2 = 0.50$ |
| **FP** | $\eta_1^{FP} = 0.01$ | $\eta_2^{FP} = 0.10$ | $\eta'^{FP}_2 = 0.10$ |
| **Ref.** | [28] | [12] | |

SI Table 3. Parameters independent of test.

| Parameter | Value | Notes |
|---|---|---|
| $I_{T1}$ | 1% (10 infections) | First (starting) threshold for testing. The testing gets started (or restarted) when the total infections in a 1000-node network is $\geq 10$. |
| $I_{T2}$ | 0.5% (5 infections) | Second (stopping) threshold for testing. The testing is stopped when the total infections in a 1000-node network is $\leq 5$. |
| $Q_D$ | 10 days | Duration of quarantine, when node is tested to be positive. Quarantine starts when the test result is obtained (after the delay between testing and obtaining of results) |

# 7. Algorithm for Simulation

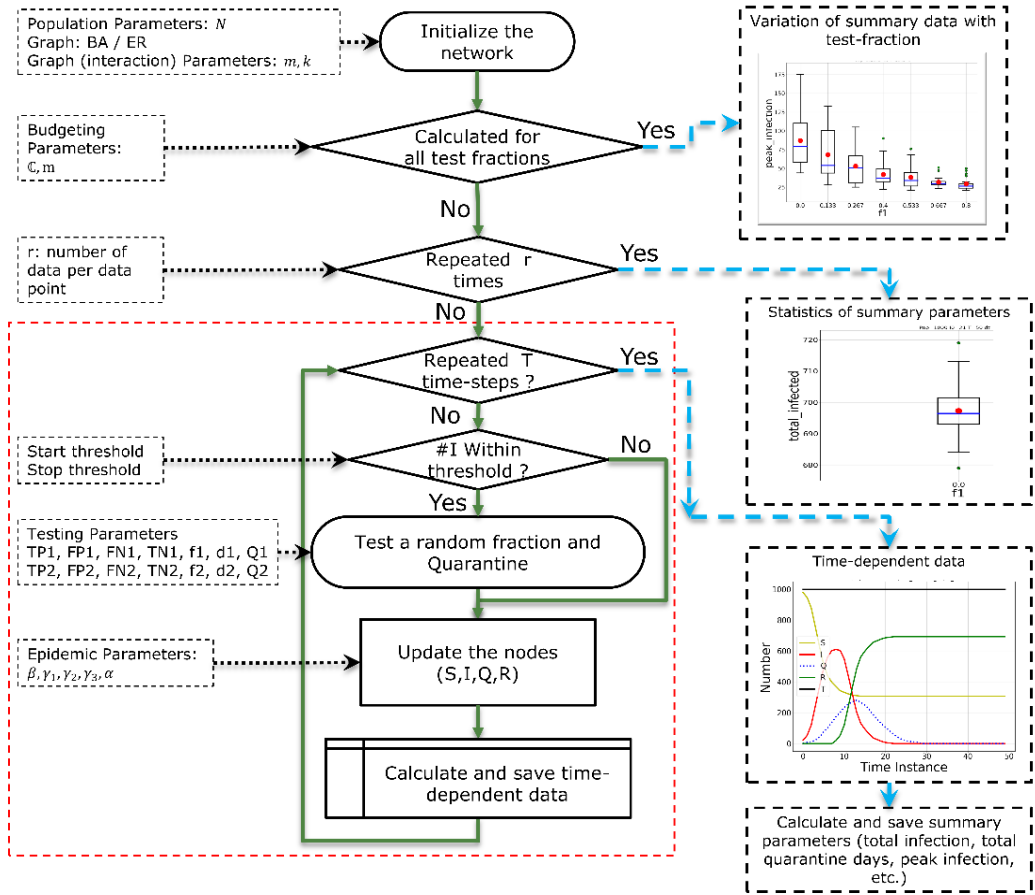

SI Figure 6

# 8. Typical Propagation of Epidemic in the Network

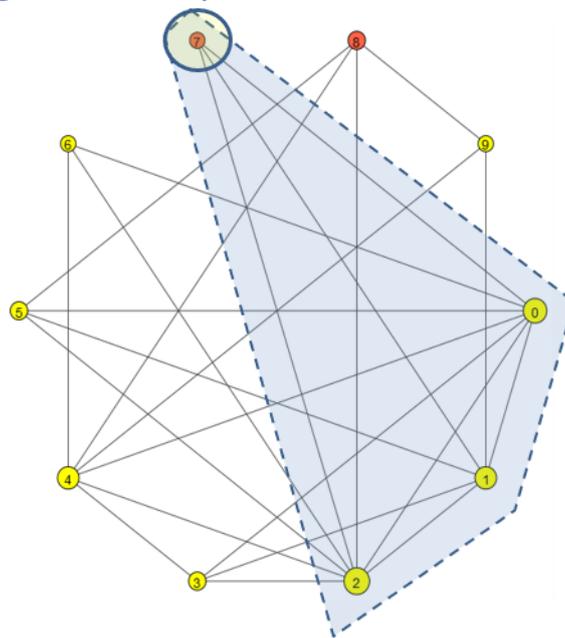

SI Figure 7

# 9. Utility Function

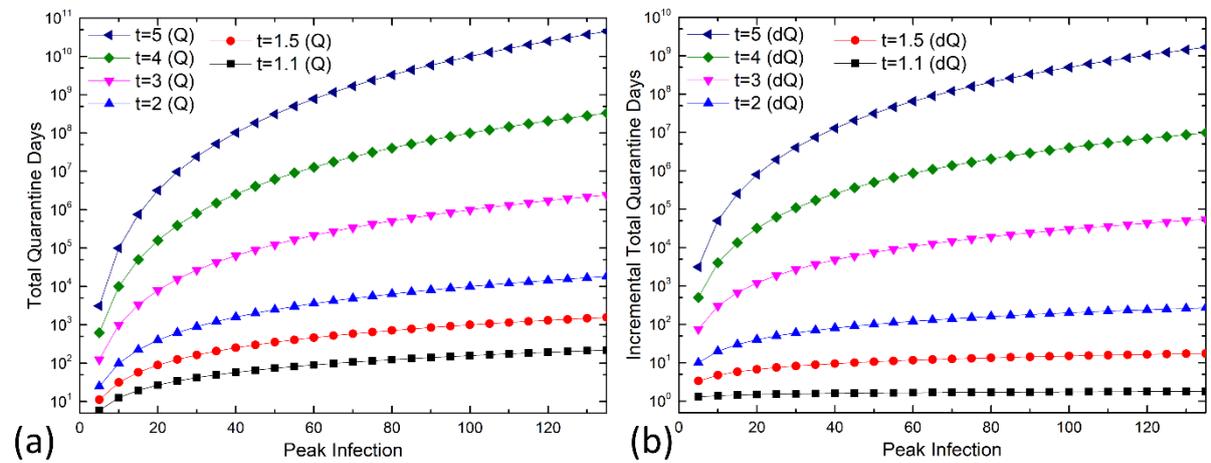

SI Figure 8

10 Quarantine days = 1 Quarantine

# 10. Effect of Test cost – complete set of graphs

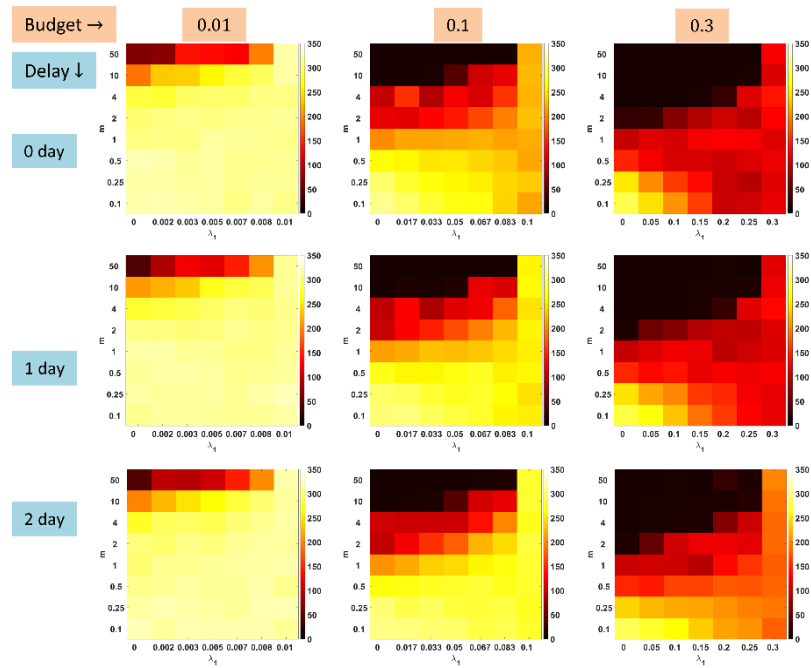

SI Figure 9. Effect of total budget (B), fractional cost of second test (m) and delay in obtaining test one result on peak infection

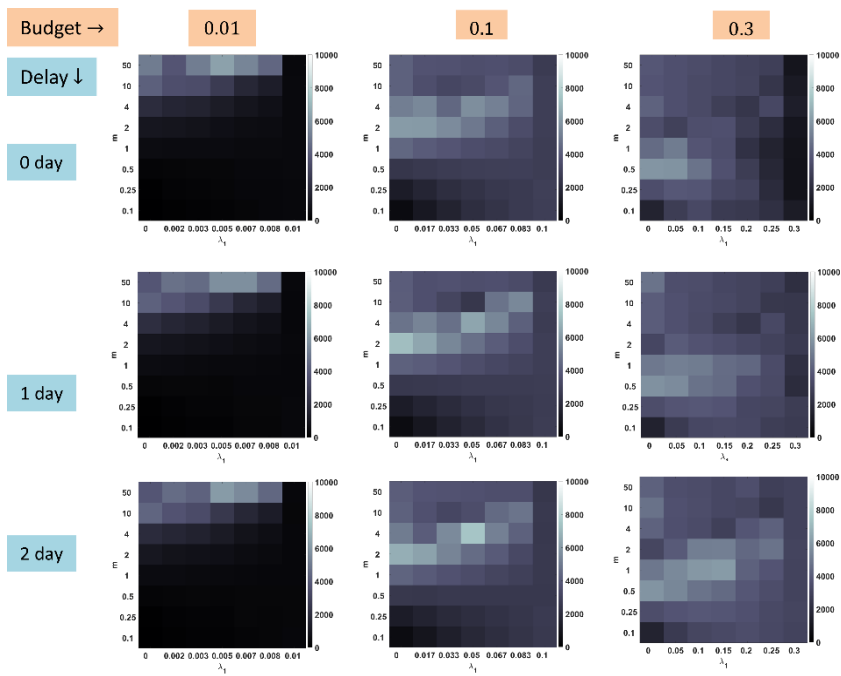

SI Figure 10. Effect of of total budget (B), fractional cost of second test (m) and delay in obtaining test one result on peak infection